\documentclass[english,showpacs,prb]{revtex4}
\usepackage{amsmath,amssymb,graphicx}

\makeatletter

\begin{document}

\title{Coulomb blockade and quantum tunnelling in an array
of metallic grains}

\author{V. Tripathi, M. Turlakov, and Y.L. Loh}


\address{Theory of Condensed Matter Group, Cavendish Laboratory, University
of Cambridge, Cambridge CB3 0HE, United Kingdom }

\begin{abstract}
We study the effects of Coulomb interaction and inter-grain
quantum tunnelling in an array of metallic grains using the
phase-functional approach
for temperatures $T$ well below the charging energy $E_{c}$ of individual
grains yet large compared to the level spacing in the grains. When
the inter-grain tunnelling conductance $g\gg1$, the conductivity $\sigma$
in $d$ dimensions decreases logarithmically with temperature 
($\sigma/\sigma_{0}\sim1-\frac{1}{\pi gd}\ln(gE_{c}/T)$)\cite{zaikin,grabert,efetov1},
while for $g\rightarrow0$, the conductivity shows simple activated behaviour ($\sigma \sim \exp(-E_c/T)$). 
We show, for bare tunnelling conductance $g \gtrsim 1$, that the
parameter 
$\gamma \equiv g(1-2/(g\pi)\ln(gE_{c}/T))$ 
determines the competition between charging and tunnelling
effects. At low enough temperatures in the regime $1\gtrsim \gamma \gg
1/\sqrt{\beta E_{c}}$,
a charge is shared among a finite number $N=\sqrt{
  (E_{c}/T)/\ln(\pi/2\gamma z)}$ of grains, and we find a soft activation
behaviour of the conductivity, 
$\sigma\sim z^{-1}\exp(-2\sqrt{(E_{c}/T)\ln(\pi/2\gamma z)})$,
where $z$ is the effective coordination number of a grain. 
\end{abstract}


\maketitle
\section{Introduction}
Coulomb effects and electron tunnelling as well as various effects of disorder
are major themes of mesoscopic physics. For a single metallic grain, all these
effects have been explored extensively. For an array of normal metal
grains, we find a peculiar
interplay of electron charging and tunnelling effects manifested by a formation
of a multi-grain charge excitation. 

The novelty of granular metal systems in comparison with disordered metals
or semiconductors with impurities arises from the presence of additional
energy scales - the grain charging energy $E_{c}$ \cite{abeles1} 
and intra-grain
energy-level spacing $\delta$. When the temperature is lowered below level
spacing $\delta$, electrons in the granular metal can propagate (diffuse
or hop over many grains) coherently just like in a disordered metal.
In the incoherent regime $T>\mbox{max }(g\delta,\delta)$, only charging
effects and nearest-neighbour hopping (in the second order of the tunnelling
matrix element) are relevant, and it is possible to formulate the
problem in terms of longitudinal electromagnetic phase (or voltage)
fluctuations on the grains\cite{efetov1,ambegaokar1}. For large inter-grain
conductance $g\gg1$, the conductivity decreases logarithmically with
temperature independent of dimensionality\cite{zaikin,grabert,efetov1},
reminiscent of many experiments\cite{chui1,simon1,gerber1}. 
The essential physics
of this result was recognised some time ago as washed out Coulomb
blockade for the quantum dot coupled to a conductive environment\cite{devoret1,girvin1}.
For intermediate conductance $g \gtrsim 1$ (and sufficiently low temperature), we derive within the
same approach a soft activation conduction ($\sigma\sim e^{-\sqrt{T_{0}/T}}$)
by a charge excitation (later referred to as a {}``puddle'') shared
between many grains due to incoherent tunnelling. In the itinerant
(or large-scale diffusion over many grains) regime at low temperatures
$T<\mbox{max }(g\delta,\delta)$, the granular metal for large conductance
$g\gg1$ is naturally described by Altshuler-Aronov theory\cite{altshuler1,beloborodov2}.

We employ the phase functional approach suggested over two decades
ago by Ambegaokar, Eckern, and Sch\"{o}n (AES) \cite{ambegaokar1}.
The original model described the tunnelling dynamics of granular 
superconductors, but nowadays this approach is 
increasingly used to study granular metals as well \cite{efetov1,beloborodov2,beloborodov1}. 
The AES action $S_{AES}$
for granular metals consists of two contributions, $S_{AES}=S_{c}+S_{t}$,
where \begin{eqnarray}
S_{c} & = & \frac{1}{2}\sum_{{\bf{i,j}}}\int_{0}^{\beta}d\tau\, C_{{\bf{ij}}}\frac{d\tilde{\phi}_{i}(\tau)}{d\tau}\frac{d\tilde{\phi}_{j}(\tau)}{d\tau}\label{charging}\end{eqnarray}
 represents charging of the grains, and \begin{equation}
S_{t}=\pi g\sum_{|{\bf{i-j}}|=a}\int_{0}^{\beta}d\tau\, d\tau'\alpha(\tau-\tau')\sin^{2}\left(\frac{\tilde{\phi}_{{\bf{ij}}}(\tau)-\tilde{\phi}_{{\bf{ij}}}(\tau')}{2}\right)\label{tunneling}\end{equation}
 represents tunnelling between neighbouring grains, $\tilde{\phi}_{{\bf{ij}}}=\tilde{\phi}_{{\bf{i}}}-\tilde{\phi}_{{\bf{j}}}$.
The kernel $\alpha(\tau)$ has the form $\alpha(\tau)=T^{2}(\mbox{Re }(\sin(\pi T\tau+i\epsilon))^{-1})^{2}$.
The fields $\{\tilde{\phi}_{{\bf{i}}}\}$ are electromagnetic
phase fluctuations on the grains related to the respective potential
fluctuations $\{ V_{{\bf{i}}}\}$ through $V_{{\bf{i}}}(\tau)=\partial_{\tau}\tilde{\phi}_{{\bf{i}}}(\tau)$.
They satisfy bosonic boundary conditions, $\tilde{\phi}_{{\bf{i}}}(\tau)=\frac{2\pi k_{{\bf{i}}}}{\beta}\tau+\phi_{{\bf{i}}}(\tau)$,
$\phi_{{\bf{i}}}(\tau)=\phi_{{\bf{i}}}(\tau+\beta)$, where
the winding number $k_{{\bf{i}}}$ is an integer, and $-\infty<\phi_{{\bf{i}}}(\tau)<\infty$.
The tunnelling conductance $g$ is related to the inter-grain hopping
amplitude $t_{{\bf{i, i+a}}}$ through $g=2\pi|t_{{\bf{i, i+a}}}|^{2}/\delta^{2}$.
Conductivity in the AES model is a second order (in hopping amplitude)
incoherent tunnelling process between neighbouring grains. The elastic
tunnelling lifetime $\tau$ of the electron on the grain is $\tau=\hbar/(g\delta)$.
The condition, defining the granularity of the material and allowing
averaging over fermionic intra-grain states, is that the tunnelling
lifetime $\tau$ is much longer than the Thouless diffusion time $l^{2}/D$
(where $D$ is intra-grain diffusion coefficient, and $l$ is a size
of a grain). 
Another relevant condition is the implicit requirement of energy
relaxation in the grains. The characteristic times associated with
these incoherent dissipative processes should be shorter than tunnelling
lifetime, consequently coherent combination of wavefunctions over
grains cannot be written.
Moreover, our diagrammatic analysis shows (see also Ref.\cite{beloborodov1})
that higher order processes ($|t_{{\bf{i, i+a}}}|^{4}$, etc.)
can be neglected when $T>g\delta$. Such a condition allows us to
neglect {}``dressing'' of the tunnelling vertex $t_{{\bf{i, i+a}}}$
by further tunnelling lines. Therefore, the phase functional approach
for granular materials can be justified if $g\delta\ll D/l^{2}$,
and the temperature is sufficiently high, $T\gg\mbox{ max}(\delta,g\delta)$.
We shall restrict our analysis to this regime. 

\section{Analysis of Model}

The AES action shows important qualitative changes in the relevance
of large phase fluctuations as the coupling $g$ is varied. Consider
the metallic phase $g\gg1$ in Eq.(\ref{tunneling}). Expanding $\sin^{2}(\tilde{\phi}_{{\bf{ij}}}(\tau)-\tilde{\phi}_{{\bf{ij}}}(\tau'))$
in a power series, we observe $\langle\tilde{\phi}_{{\bf{ij}}}^{2}\rangle\sim g^{-1}$,
thus inter-grain phase fluctuations are Gaussian, and suppressed. 
The charge on an individual grain is not a well
defined quantity, rather it is shared by the entire system. As $g$
is progressively reduced, the phase fluctuations in Eq.(\ref{tunneling})
increase until finally one needs to take into account non-zero winding
numbers $k_{{\bf{i}}}\neq0$. In the extreme limit of $g\rightarrow0$,
the AES model describes a system of weakly coupled capacitors. The
phase fluctuations are large, however the charge on an individual
grain is well-defined. Conduction now involves exciting a charge which
results in an activated temperature dependence of conductivity ($\sigma\sim g\exp(-\beta E_{c})$).
Such considerations lead us to examine whether for intermediate
coupling between the grains, charge could be shared by a finite number
of grains. This would be an intermediate situation between the extreme
cases discussed above. In the remaining part of this letter, we choose
a diagonal capacitance matrix in Eq.(\ref{charging}), $C_{{\bf{ij}}}\approx\frac{1}{2E_{c}}\delta_{{\bf{ij}}}$,
to keep our analysis simple. 

We describe now the physical picture for the soft activation phase
which emerges from our analysis. Putting a single electron on an isolated
grain costs $E_{c}$, while incoherent tunnelling enables the charge to
be shared between two or more grains. We show below that
the parameter $\gamma \equiv g(1-2/(\pi g)\ln(g\beta E_{c}))$
controls the suppression of winding numbers, and determines the degree
of charge delocalisation. When $\gamma \gtrsim 1$, the charge is
effectively delocalised over the entire system (the charging energy is
exponentially suppressed). For $\gamma \lesssim 1$, a unit of charge
(electron) is shared among a finite number of grains. For simplicity
we consider two grains, and compare statistical weights associated
with the charge localised on any single grain $P_{1}\sim\exp(-\beta
E_{c})$, and the charge shared between the two grains, 
$P_{2}\sim \gamma\exp(-\beta E_{c}/2)$
($\beta=1/T$). Observe that the charging energy is halved
upon hybridisation\cite{hybrid1}. Since a charge is shared
classically (incoherently) between two grains, it is equivalent to
equal average voltage on the grains, and thus two capacitors connected
in parallel. The total capacitance is doubled, and the charging energy
is halved.  
If $\gamma < \exp(-\beta E_{c}/2)$, the
charge is unlikely to be shared between the two grains. If on the
other hand, $\gamma > \exp(-\beta E_{c}/2)$, 
the electron is more likely to live
on both the grains. 
Thus the two-grain hybridisation ``puddle'' 
optimising the charging and tunnelling
energies is formed.
The optimum number $N_{*}$
of hybridised grains sharing a single charge is determined by maximising
$P_{N}\sim \gamma^{N-1}\exp(-\beta E_{c}/N)$, which gives $N_{*}\sim\sqrt{\beta E_{c}/\ln(\gamma^{-1})}$,
hence 
$\sigma\propto gP_{N_{*}}\sim\exp(-2\sqrt{\beta E_{c}\ln(\gamma^{-1})})$.
This in essence is our main result.

To calculate the conductivity, we use Kubo's formula\cite{efetov1},
\begin{eqnarray}
\sigma(\omega) & = & \frac{ia^{2-d}}{\omega}\pi g\int_{0}^{\beta}d\tau\,\alpha(\tau)\,(1-e^{i\Omega_{n}\tau})\times\nonumber \\
 &  & \times\left\langle \cos(\tilde{\phi}_{{\bf{i,i+a}}}(\tau)-\tilde{\phi}_{{\bf{i,i+a}}}(0)\right\rangle \big|_{\Omega_{n}\rightarrow-i\omega+\epsilon},\label{kubo}\end{eqnarray}
 where $\Omega_{n}=\frac{2\pi}{\beta}n$. Also of interest is the
tunnelling density of states $\nu_{{\bf{i}}}(\varepsilon)$ into
the grain ${\bf{i}}$: \begin{eqnarray}
\frac{\nu_{{\bf{i}}}(\varepsilon)}{\nu_{0}T} & = & \mbox{Im}\left[\int d\tau\,\frac{e^{i\varepsilon_{n}\tau}}{\sin(\pi\tau T)}\tilde{\Pi}_{{\bf{i}}}(\tau)\bigg|_{\varepsilon_{n}\rightarrow-i\varepsilon+\delta}\right],\label{tds}\end{eqnarray}
 where $\tilde{\Pi}_{{\bf{i}}}=\langle\exp(-i(\tilde{\phi}_{{\bf{i}}}(\tau)-\tilde{\phi}_{{\bf{i}}}(0)))\rangle$,
and $\varepsilon_{n}=\frac{2\pi}{\beta}(n+\frac{1}{2})$. 

At this stage we are in a position to understand qualitatively the
logarithmic temperature dependence of the conductivity for $g\gg1$ 
(derived in Ref.\cite{zaikin,grabert,efetov1}).
Since in this regime phase fluctuations are small, we set all $k_{{\bf i}}=0$
and expand $S_{t}$ to quartic order in $\phi_{{\bf ij}}$. Denoting
the Gaussian part of the resulting AES action as the {}``free''
action, and considering the quartic bit as {}``interaction'',
one finds that the interaction renormalises the tunnelling conductance
as \cite{efetov1} \begin{eqnarray}
g_{ren}(\tau-\tau') & \approx & g(1-\langle\phi_{{\bf ij}}(\tau)\phi_{{\bf ij}}(\tau')\rangle),\label{grenormal1}\end{eqnarray}
 thus in $d$ dimensions,
$g_{ren}(\beta)\approx g\left(1-\frac{1}{\pi gd}\ln(g\beta E_{c})\right)$.
One infers from Eq.(\ref{grenormal1})
a similar temperature dependence for the conductivity $\sigma$ as 
it is proportional to the effective
tunnelling conductance $g_{ren}$. This result suggests that when
temperature $T$ falls below $T_{0}=E_{c}e^{-\pi d(g-1)}$, a transition
or crossover into an insulating phase might be expected. 
Note that the parameter $\gamma$ becomes smaller than one at temperature
$T_1=E_c e^{-\pi (g-1)/2}$, which is parametrically much larger than
$T_0$. We show that $T_1$ marks the onset of soft activation
behaviour.

Consider the AES model, Eq.(\ref{charging}) and Eq.(\ref{tunneling}).
At low temperatures and for $g \gtrsim 1$, phase fluctuations on the grains
could be large and non-Gaussian, so we expand the action about the 
finite winding number phase changes $(2\pi/\beta)k_{{\bf{i}}}\tau$
and residual fluctuations $\phi_{{\bf{i}}}$\cite{falci}:
\begin{eqnarray}
&S&[\{ k_{{\bf{i}}}\};\{\phi_{{\bf{i}}}(\omega_{n})\}]  = 
 \frac{(2\pi)^{2}}{4\beta E_{c}}\sum_{{\bf{i}}}k_{{\bf{i}}}^{2}
  +  \pi g\sum_{|{\bf{i-j}}|=a}|k_{{\bf
 ij}}|+ \nonumber \\
&+& \frac{\beta}{4E_{c}}\sum_{{\bf{i}},n}\omega_{n}^{2}\,\phi_{{\bf{i}}}(\omega_{n})\phi_{{\bf{i}}}(-\omega_{n})
   +\frac{\beta
  g}{2}\sum_{|{\bf{i-j}}|=a}\left(|\omega_{n+k_{{\bf{ij}}}}|+|\omega_{n-k_{{\bf{ij}}}}|-2|\omega_{k_{{\bf ij}}}|\right)\times \nonumber \\
&\times& \phi_{{\bf{ij}}}(\omega_{n})\phi_{{\bf{ij}}}(-\omega_{n})+O(\phi^{4}),\label{AES2}\end{eqnarray}
 where $\phi_{{\bf{i}}}(\tau)=\sum_{n}\phi_{{\bf{i}}}(\omega_{n})\exp(i\omega_{n}\tau)$.
Since the bare conductance is large, $g\gtrsim 1$, an expansion to quadratic
 order in the residual fluctuations is justified.
The first two terms of Eq.(\ref{AES2}) arise from finite-winding
number (non-Gaussian) fluctuations, and directly lead to quantisation
of charge. The remaining terms in Eq.(\ref{AES2}) arise from perturbation
about the winding numbers. 
The competition of single-grain charging and hybridisation at low
 enough temperatures
can be seen in the partition function 
$Z_{2}=\int D\phi_{1}D\phi_{2}\sum_{k_{1},k_{2}}\exp(-S[k,\phi])$
of a simple two-grain system. 
Since the tunnelling term depends only on the phase difference between
 the two grains, we make a transformation to average phase
 $\phi_{av}=(\phi_{1}+\phi_{2})/2$, and relative phase $\phi =
 (\phi_{1}-\phi_{2})$. Integrating out the relative phase gives a
 winding number dependent determinant. This we normalise against the
 determinant with no winding numbers:
\begin{eqnarray}
\frac{\mbox{Det}_{\phi}[k_{12}=0]}{\mbox{Det}_{\phi}[k_{12}]}
   = 
\frac{\prod_{n=1}^{\infty}\left[\frac{\pi n^{2}}{4\beta
 E_{c}}+g|n|\right]}{\prod_{n=1}^{\infty}\left[\frac{\pi n^{2}}{4\beta
 E_{c}}+\frac{g}{2}(|n+k_{12}|+|n-k_{12}|-2|k_{12}|)\right]} \nonumber
 \\
= \prod_{n=1}^{|k_{12}|}\left[1+\frac{4g\beta
 E_{c}}{n\pi}\right]\prod_{n=|k_{12}|+1}^{\infty}\left[\frac{\frac{\pi
 n^{2}}{4\beta E_{c}}+gn}{\frac{\pi n^{2}}{4\beta E_{c}} + 
g(n-|k_{12}|)}\right].
\label{detphi1}
\end{eqnarray}
For $k_{12}\ll g\beta E_{c}$ (a large $g$ suppresses large
values of $k_{12}$), the determinant in Eq.(\ref{detphi1}) can be
simplified as
\begin{eqnarray}
\frac{\mbox{Det}_{\phi}[k_{12}=0]}{\mbox{Det}_{\phi}[k_{12}]}
 & \sim & \frac{1}{|k_{12}|!}\left(\frac{4g\beta
  E_{c}}{\pi}\right)^{|k_{12}|}
\prod_{n=|k_{12}|+1}^{4g\beta E_{c}/\pi}\frac{n}{n-|k_{12}|}\nonumber
 \\
 & = & \frac{(4g\beta E_{c}/\pi)^{|k_{12}|}(4g\beta E_{c}/\pi)!}{(4g\beta
 E_{c}/\pi -|k_{12}|)!\,(|k_{12}|!)^{2}} \nonumber \\
 & \sim & \exp\left[2|k_{12}|\ln\left(\frac{4ge\beta E_{c}}{\pi|k_{12}|}\right)\right],
\end{eqnarray} 
where we used Stirling's formula for the factorials. 
This result implies that relative phase fluctuations enhance the tendency
for phase slips between neighbouring grains. The effective action for
the two grain system takes the form
\begin{eqnarray}
S[{k},\phi_{av}] & = & \frac{(2\pi)^{2}}{4\beta
  E_{c}}\sum_{i}k_{i}^{2} + \frac{2\beta}{4
  E_{c}}\sum_{n}\omega_{n}^{2}\phi_{av}(\omega_{n})\phi_{av}(-\omega_{n})
  + \nonumber \\
   &  & + \pi g|k_{12}|\left[1-\frac{2}{\pi g}\ln\left(\frac{4ge\beta
  E_{c}}{\pi|k_{12}|}\right)\right] \nonumber \\
  & = & \frac{(2\pi)^{2}}{4\beta
  E_{c}}\sum_{i}k_{i}^{2} + \pi \gamma|k_{12}| + \nonumber \\
   &  & +  \frac{2\beta}{4
  E_{c}}\sum_{n}\omega_{n}^{2}\phi_{av}(\omega_{n})\phi_{av}(-\omega_{n}), 
\label{twograinaction}
\end{eqnarray}
where the effective parameter
\begin{equation}
\gamma = g\left[1-\frac{2}{\pi g}\ln\left(\frac{4ge\beta E_{c}}{\pi|k_{12}|}\right)\right],
\label{effectivegamma}
\end{equation}
is smaller than the renormalised tunnelling $g_{ren}=g(1-1/(\pi
dg)\ln(g\beta E_{c}))$ that represents the conductance at large bare
coupling $g$ at not too low temperatures. The main interest of this
paper is the regime $\gamma \lesssim 1$. If $1\gtrsim \gamma \gg 1/(\beta
E_{c})$, most contribution to single charge excitations in the 
two-grain partition function comes from low
winding number difference ($|k_{12}|\sim O(1)$) between the grains. On the other hand, when
$\gamma \lesssim 1/(\beta E_{c})$, large winding number differences of the order of
$\beta E_{c}$ become important. Therefore in this regime, the
charge is localised on either of the grains, with large charging
energy of order $E_{c}$. 

It is important to obtain the charge representation by summing 
over the winding numbers with the use of the Poisson formula 
\begin{eqnarray}
Z_{2} & = & \sum_{\{
  q_{i}\}}\int_{-\infty}^{\infty}D\phi_{av}\frac{dx_{1}}{2\pi}\frac{dx_{2}}{2\pi}e^{i2\pi\sum_{i}q_{i}x_{i}-S[x,\phi_{av}]}.
\label{poisson1}\end{eqnarray} 
 Integrating out $x_{2}$ will yield the effective environment around
$x_{1}$: \begin{eqnarray}
Z_{2}\approx\sum_{q_{1},q_{2}}\int dx_{1}\int D\phi_{av}\, 
e^{-\frac{2\beta}{4E_{c}}\sum_{n}\omega_{n}^{2}\,\phi_{av}(\omega_{n})\phi_{av}(-\omega_{n})}\times\nonumber \\
\times\left(\sqrt{\frac{\beta E_{c}}{\pi}}\, e^{\frac{\gamma^{2}}{4\beta E_{c}}-\beta E_{c}q_{2}^{2}-\frac{(2\pi)^{2}}{4\beta E_{c}}x_{1}^{2}+i2\pi q_{1}x_{1}}\right.\times\nonumber \\
\times\left[\Theta(x_{1})e^{i\gamma\beta E_{c}q_{2}-\pi \gamma
    x_{1}}+\Theta(-x_{1})e^{-i\gamma\beta E_{c}q_{2}+\pi \gamma x_{1}}\right]+\nonumber \\
+\left.\frac{\gamma/2\pi}{(\gamma/2)^{2}+(q_{2}+ix_{1}\pi/\beta E_{c})^{2}}e^{-\frac{2(2\pi)^{2}}{4\beta E_{c}}x_{1}^{2}+i2\pi x_{1}(q_{1}+q_{2})}\right).\label{env1}\end{eqnarray}
 There are two qualitatively distinct contributions in Eq.(\ref{env1}).
The first two terms represent isolated charging of grain $2$. If
$\gamma$ were vanishingly small, this would be the only contribution.
The last term represents hybridisation of the two grains because of
quantum tunnelling; the total charge $q_{1}+q_{2}$ is shared between
the grains, and the charging energy $E_{c}$ is reduced to $E_{c}/2$.
Finally, integration over $x_{1}$ gives the relative weights of the
two processes in the partition function as $P_{1}\approx\exp(-q_{1}^{2}\beta E_{c})$
for isolated charging, and $P_{2}\approx\frac{2\gamma}{\pi}\exp(-(q_{1}+q_{2})^{2}\beta E_{c}/2)$
for charging of the hybridised grains. 

The treatment so far considers residual relative phase fluctuations
only to gaussian order. In the Appendix, we present the results of
path integral Monte Carlo calculations to support our basic idea of
charge sharing over two grains, even at low temperatures, where
non-gaussian fluctuations are important. We also 
confirm that the temperature dependence of $\gamma$ agrees
well with Eq.(\ref{effectivegamma}) at not too low temperatures.
As the temperature is decreased further, non-gaussian fluctuations become
important. The numerical calculations show that our physical
picture, that a competition of charging and
tunnelling effects determines whether the charge is shared
between the two grains (with charging energy $E_{c}/2$) or localised
on a single grain, still remains valid. More precisely, we find that the probability of
sharing a charge between two grains (here $\gamma$) is an algebraically, and not
exponentially, small function of temperature\cite{mycomment}. 
 
Consider now the case of $N$ connected grains. 
Formally, it is simple enough to demonstrate annexation of a single
grain into an $N$-site puddle. The proof is by induction. Suppose that
an $N$-site puddle already exists (with statistical weight $\propto
\gamma^{N}$). 
Integration over a string of $N-1$ contiguous neighbours of a grain
${\bf{i}}$ similarly gives a puddle of size $N$ with charging
energy $E_{c}/N$, and a weight
$P_{N}\approx(\frac{2\gamma}{\pi})^{N-1}\exp(-(q_{1}+\cdots+q_{n})^{2}\beta
E_{c}/N)$.
Such an expansion in $\gamma$ only makes sense
if $2\gamma/\pi <1$. For $2\gamma/\pi > 1$, optimum size 
of the puddle is divergent.
Consider the action of a single grain coupled to this
puddle:
\begin{eqnarray}
S[k,\phi]  =  \frac{(2\pi)^{2}Nk_{N}^{2}}{4\beta E_{c}}
   + \frac{(2\pi)^{2}k_{N+1}^{2}}{4\beta E_{c}} + \pi g |k_{N,N+1}| +
   \nonumber \\
   +
   \frac{N\beta}{4E_{c}}\sum_{n}\omega_{n}^{2}|\phi_{N}(\omega_{n})|^{2}
   +
   \frac{\beta}{4E_{c}}\sum_{n}\omega_{n}^{2}|\phi_{N+1}(\omega_{n})|^{2}
   + \nonumber \\
    + \frac{\beta
   g}{2}\sum_{n}(|\omega_{n+k_{N,N+1}}|+|\omega_{n-k_{N,N+1}}|-2|k_{N,N+1}|)\times
   \nonumber \\
 \times |\phi_{N,N+1}(\omega_{n})|^{2}.
\label{intermediateNaction}
\end{eqnarray}
In terms of the centre of mass coordinate
\begin{equation}
\phi_{av}= \frac{N\phi_{N}+\phi_{N+1}}{N+1},
\end{equation}
and relative coordinate 
\begin{equation}
\phi = \phi_{N}-\phi_{N+1},
\end{equation}
the action takes the form
\begin{eqnarray}
S[k,\phi]  =  \frac{(2\pi)^{2}Nk_{N}^{2}}{4\beta E_{c}}
   + \frac{(2\pi)^{2}k_{N+1}^{2}}{4\beta E_{c}} + \pi g |k_{N,N+1}| +
   \nonumber \\
   +
   \frac{\beta}{4E_{c}}\sum_{n}\omega_{n}^{2}\left[(N+1)|\phi_{av}(\omega_{n})|^{2}
   + \frac{N}{N+1}|\phi(\omega_{n})|^{2}\right]
   + \nonumber \\
    + \frac{\beta
   g}{2}\sum_{n}(|\omega_{n+k_{N,N+1}}|+|\omega_{n-k_{N,N+1}}|-2|k_{N,N+1}|)\times
   \nonumber \\
 \times |\phi(\omega_{n})|^{2}.\qquad
\label{Ngrainaction}
\end{eqnarray}
Integrating out the relative phase renormalises the bare coupling $g$
in a manner similar to that in Eq.(\ref{effectivegamma}),
\begin{equation}
\gamma_{N,N+1} = g\left[1-\frac{2}{\pi g}\ln\left(\frac{2eg\beta
    E_{c}}{\pi |k_{N,N+1}|}\frac{N}{N+1}\right)\right].
\label{Neffectiveg}
\end{equation}
Note that the relevant $\gamma_{N,N+1}$ determining annexation of a
single grain into an N-site puddle is not too different from $\gamma$
for a two grain system obtained in Eq.(\ref{effectivegamma}).
Accordingly, the condition for suppression of large winding number
difference changes from $\gamma \gg 1/(\beta E_{c})$ for two grains to
the condition $\gamma \gg N/(\beta E_{c})$ for $N$ grains.
Performing the summation over $k_{N}$ and $k_{N+1}$ in
Eq.(\ref{Ngrainaction}) using the Poisson summation formula again
yields two terms that correspond to separate charging of the puddle
and grain, and charging of the larger (N+1)-site puddle. The criterion
for annexation is 
\begin{equation}
\frac{2\gamma_{N,N+1}}{\pi}\exp[-\beta E_{c}/(N+1)] > \exp[-\beta
  E_{c}/N].
\end{equation}  

So far we have obtained the effective environment of a site ${\bf{i}}$
by integrating out a sequence of $N-1$ contiguous
neighbours. Integrating over such `strings' is somewhat different from
the actual requirement that one should consider an arbitrary puddle
with $N$ sites, and integrate over all $N$ phases. Since 
the number of bonds exceeds the number of
sites in two and three dimensions, it would be incorrect to consider
the phase differences between bonds as independent variables. The
maximum number of independent phase differences in a puddle of $N$
sites is $N-1$. Starting from an arbitrary site in the puddle, a
non self-intersecting string of $N-1$ bonds spans all $N$ sites. The
string, however, is not unique, hence in the partition function
$Z_{N}$ for $N$ coupled sites, one must consider all possible
self-avoiding string configurations of $N-1$ links.  
From the theory of self-avoiding random
walks\cite{degennes}, it is known that the degeneracy ${\mathcal{N}}$
of such configurations is \begin{eqnarray}
{\mathcal{N}}(N) & \sim & \left\{ \begin{array}{c}
(N-1)^{1/6}\tilde{z}_{3}^{N-1},\quad d=3\\
(N-1)^{1/3}\tilde{z}_{2}^{N-1},\quad d=2\end{array}\right.,\label{degen}\end{eqnarray}
 where $\tilde{z}_{d}$ is an effective coordination number that depends
on the dimensionality and the arrangement of grains. For a simple
cubic lattice in three dimensions, $\tilde{z}_{3}=4.68$, slightly
less than $6$, which is the actual coordination number. Thus the
contribution of an $N$-site puddle to the partition function,
say in three dimensions, is \begin{eqnarray}
Z_{N}\approx(N-1)^{1/6}\sqrt{\frac{\beta E_{c}}{\pi N}}\left(\frac{2\gamma\tilde{z}_{3}}{\pi}\right)^{N-1}\exp\left(-\frac{\beta E_{c}}{N}q_{N}^{2}\right)\times\nonumber \\
\times\int D\phi\exp\left(-\frac{N\beta}{4E_{c}}\sum_{n}\omega_{n}^{2}\phi(\omega_{n})\phi(-\omega_{n})\right).\qquad\label{Zn}\end{eqnarray}
 The optimum size of the puddle is reached when $N=N_{*}\approx\sqrt{\frac{\beta E_{c}q_{N_{*}}^{2}}{\ln(\pi/2\tilde{z}_{3}\gamma)}}$,
and the dominant contribution to the partition function is \begin{eqnarray}
Z_{N_{*}} & \approx & \frac{\pi}{2\gamma\tilde{z}_{3}}\exp\left(-2\sqrt{\beta E_{c}q_{N_{*}}^{2}\ln(\pi/2\tilde{z}_{3}\gamma)}\right).\label{zn2}\end{eqnarray}
This result is valid under the condition $1\gtrsim \gamma \gg
1/\sqrt{\beta E_{c}}$. 

We now have the necessary ingredients for calculating the conductivity
$\sigma$ and tunnelling density of states $\nu_{{\bf{{i}}}}$
from Eq.(\ref{kubo}) and Eq.(\ref{tds}). Calculation of the conductivity
$\sigma$ using Eq.(\ref{kubo}) requires evaluation of a two-point
phase correlation function $\tilde{\Pi}_{{\bf{i, i+a}}}$, \begin{eqnarray}
\tilde{\Pi}_{{\bf{i, i+a}}} & = & \left\langle \exp\left(-i\left(\tilde{\phi}_{{\bf{i, i+a}}}(\tau)-\tilde{\phi}_{{\bf{i, i+a}}}(0)\right)\right)\right\rangle .\label{phasecorrel3}\end{eqnarray}
 The two points ${\bf{i}}$ and ${\bf{i+a}}$ should be chosen
to lie in different puddles, for if they lie within the same puddle,
$\tilde{\Pi}$ would simply describe fluctuation of charge distribution
inside a puddle; this contributes little to the conductivity $\sigma$.
This simplifies evaluation of the two-point phase correlation function
to a product of two one-point phase correlation functions, $\tilde{\Pi}_{{\bf{i, i+a}}}\approx\langle\exp(-i(\tilde{\phi}_{{\bf{i}}}(\tau)-\tilde{\phi}_{{\bf{i}}}(0)))\rangle\langle\exp(i(\tilde{\phi}_{{\bf{i+a}}}(\tau)-\tilde{\phi}_{{\bf{i+a}}}(0)))\rangle$.
The averaging in Eq.(\ref{phasecorrel3}) should be performed over
winding numbers $\{ k_{{\bf{i}}}\}$ as well as the phase fluctuations
$\{\phi_{{\bf{i}}}\}$.
The AES action in Eq.(\ref{intermediateNaction}) after integrating
over the relative residual phase fluctuations
then takes the form \begin{eqnarray}
S[\{
  k_{{\bf{i}}}\};\{\phi_{{\bf{i}}}(\omega_{n})\}]=\frac{(2\pi)^{2}}{4\beta E_{c}}\sum_{{\bf{i}}}N_{{\bf i}}k_{{\bf{i}}}^{2}+\pi \gamma\sum_{|{\bf{i-j}}|=a}|k_{{\bf{i}}}-k_{{\bf{j}}}|+\nonumber \\
+\frac{\beta}{4E_{c}}\sum_{n}N_{{\bf i}}\omega_{n}^{2}\,\phi_{{\bf{i}}}(\omega_{n})\phi_{{\bf{i}}}(-\omega_{n})+\cdots.\qquad\qquad\label{AES3}\end{eqnarray}
 Upon performing the average, we obtain an expansion in increasing
puddle size: \begin{eqnarray}
\tilde{\Pi}_{{\bf{i, i+a}}}\approx\sum_{\{ N_{{\bf{i}}}\}}\left(\frac{2\gamma\tilde{z}_{d}}{\pi}\right)^{N_{{\bf{i}}}+N_{{\bf{i+a}}}-2}e^{-E_{c}\tau\left(\frac{1}{N_{{\bf{i}}}}+\frac{1}{N_{{\bf{i+a}}}}\right)}\times\nonumber \\
\times\sum_{\{ q_{N}\}}e^{2\tau E_{c}\left(\frac{q_{N_{{\bf{i}}}}}{N_{{\bf{i}}}}-\frac{q_{N_{{\bf{i+a}}}}}{N_{{\bf{i+a}}}}\right)-\beta E_{c}\left(\frac{q_{N_{{\bf{i}}}}^{2}}{N_{{\bf{i}}}}+\frac{q_{N_{{\bf{i+a}}}}^{2}}{N_{{\bf{i+a}}}}\right)}.\label{phasecorrel5}\end{eqnarray}
 To calculate the conductivity $\sigma$ given by Eq.(\ref{kubo}),
we make the analytic continuation $\Omega_{n}\rightarrow-i\omega+\epsilon$,
and deform\cite{efetov1} the contour of integration in the following
manner: $(0,\beta)\rightarrow(0,i\infty)+(i\infty,i\infty+\beta)+(i\infty+\beta,\beta)$.
For d.c. conductivity, we expand Eq.(\ref{kubo}) for small $\omega$,
and take the limit $\omega\rightarrow0$. Performing the integration
yields the conductivity \begin{eqnarray}
\sigma & \sim & 2ga^{2-d}\sum_{N_{{\bf{i}}},N_{{\bf{i+a}}}}\sum_{q_{N_{{\bf{i}}}},q_{N_{{\bf{i+a}}}}}\left(\frac{2\gamma\tilde{z}_{d}}{\pi}\right)^{N_{{\bf{i}}}+N_{{\bf{i+a}}}-2}\times\nonumber \\
 &  & \!\!\!\!\!\!\!\!\!\!\!\times\exp\left(-\beta E_{c}\left(\frac{(q_{N_{{\bf{i}}}}-1)^{2}}{N_{{\bf{i}}}}+\frac{(q_{N_{{\bf{i+a}}}}+1)^{2}}{N_{{\bf{i+a}}}}\right)\right).\label{conductivity3}\end{eqnarray}
 Most of the contribution to Eq.(\ref{conductivity3}) comes from
two single-charge configurations $(q_{N_{{\bf{i}}}},q_{N_{{\bf{i+a}}}})=(1,0),\mbox{ or }(0,-1)$.
In the former configuration, conductivity is dominated by $(N_{{\bf{i}}},N_{{\bf{i+a}}})=(1,N_{*})$,
while in the latter configuration, conductivity is dominated by $(N_{{\bf{i}}},N_{{\bf{i+a}}})=(N_{*},1)$,
and $N=N_{*}\approx\sqrt{\frac{\beta E_{c}}{\ln(\pi/2\tilde{z}_{d}\gamma)}}$
as usual. The result is \begin{eqnarray}
\sigma & \sim & \frac{1}{\tilde{z}_{d}}a^{2-d}\exp(-2\sqrt{\beta E_{c}\ln(\pi/2\gamma\tilde{z}_{d})}).\label{conductivity4}\end{eqnarray}

We can similarly obtain the tunnelling density of states: \begin{equation}
\nu(\varepsilon)  \approx  \frac{\pi\nu_{0}}{2\gamma\tilde{z}_{d}}\cosh(\beta\varepsilon)\exp(-2\sqrt{\beta E_{c}\ln(\pi/2\gamma\tilde{z}_{d})}).\label{tds3}\end{equation}

\section{Conclusion}

We propose that our simple model of a regular array may explain soft activation
behaviour observed in real granular metals \cite{abeles1,chui1,simon1,gerber1}.
 In real granular metals, inter-grain
tunnelling may vary strongly between grains, but even in the presence
of disorder, our physical mechanism could be applicable. Firstly, for weak
disorder, suppose the inter-grain coupling for the $i^{th}$ tunnelling link
has a distribution 
$\gamma_{i}=\gamma^{1+\epsilon_{i}}.$
Then for an $N$-site puddle, since the $\epsilon_{i}$ are random, 
the tunnelling term $\prod_{i}\gamma_{i}\sim
\gamma^{N+\sum_{i} \epsilon_{i}} \sim \gamma^{N}$
is not seriously modified, and our conclusions hold.
Secondly, as discussed in the context of granular superconductors\cite{jaeger},
theoretical calculations based on regular Josephson arrays seem to
be relevant. The reason is that even for a wide distribution of
couplings, only a narrow range of couplings is relevant, since 
(a) the extremely weak links can
effectively be disregarded and (b) for links that are much
stronger than average, one can approximate the connected grains as
one single grain. 
While the tunnelling probability changes exponentially
with length, the charging energy changes only linearly, so the variation
of charging energies is relatively small. The system then effectively
consists of such renormalised 'grains' linked by tunnelling of similar
magnitude. 
Thirdly, if conduction occurs through a few 1D paths, our result,
being dimensionality independent, still applies.
The observations\cite{simon1,gerber1},
according to our picture, 
are robust even upon application
of strong magnetic fields ($\gtrsim10$ tesla) and are independent
of dimensionality\cite{gerber1}. Nevertheless, further work needs to be done
to understand properly granular metals with strong variation in inter-grain
couplings. 
 
The AES approach, we use, views conduction as a Fermi Golden-Rule
type incoherent tunnelling process. The obvious difference between
our picture of soft activation and the
Efros-Shklovskii\cite{efros1}(ES)
theory, which also gives a similar temperature dependence of
conductivity, is the on-site charging energy cost $E_{c}$ in our model and
lack of thereof in ES theory. Furthermore, the mutual interaction
of charges (and excitonic effects) on widely separated grains plays
no significant role in our analysis unlike in ES theory. Since the
soft activation mechanism involves only nearest-neighbour hopping in
comparison with long-distance variable range hopping, the magnetoresistance
of soft activation here is expected to be very weak, which is 
consistent with experiments\cite{gerber1}. 
Another possibility\cite{abeles1}
considered in the literature suggests that the observed soft activation
could be an artifact of a special distribution of grain sizes. Such
a hope is belied by observation \cite{simon1} of the same soft activation
in samples with a very narrow distribution of grain sizes. Also if
we accept the conduction process as proceeding through tunnelling of
charge between neighbouring grains, there would be little likelihood
of finding the percolation paths in the wide range of temperatures
through appropriately sized grains, should they exist, as neighbours.

The relevance of our results as well as
Refs.\cite{efetov1,beloborodov2}
should be explored
beyond carefully prepared granular arrays. Recently, a logarithmic
temperature dependence of conductivity in strong magnetic fields\cite{ando1,boebinger1}
and granular (or domain) structure\cite{lang1} has been observed
in certain underdoped cuprates. The insulating phase (even more underdoped)
in the same materials exhibited the soft activation behaviour\cite{cheong1},
which may be due to the mechanism proposed in this paper. 

In conclusion, our analysis of transport in granular arrays at not
too low temperatures $T\gg\mbox{max }(\delta,g\delta)$ in the framework
of the AES approach shows that the transitions from a logarithmic
temperature dependence of conductivity for strong inter-grain coupling
($g\gg 1$) to the soft activation behaviour for intermediate coupling
($g \gtrsim 1,\,\, 1\gtrsim \gamma \gg 1/\sqrt{\beta E_{c}}$) and further to the hard activation
behaviour for weak coupling $g\ll 1$ can be understood
as arising from the competition between Coulomb blockade and tunnelling.
This analysis is strictly valid for regular arrays and may be
considered for experimental systems\cite{abeles1,chui1,simon1,gerber1}. 

\section*{Acknowledgements}
We thank 
I.L. Aleiner, K.B.Efetov, P.B.Littlewood, and B.I.Shklovskii for valuable
and critical discussions. Acknowledgement of discussions should not necessarily
imply the endorsement of our results. 
V.T. and Y.L.L. are grateful to Trinity College, Cambridge for support. 

\section*{Appendix}
\textit{Note added in proof:}
The main idea of our paper is that charge sharing among several grains
occurs at certain low temperatures, and the probability of sharing a
charge between two grains is not an exponentially small function of temperature. 
In this Appendix we discuss various arguments (in addition
to the calculation in the main text) in support
of this main idea, which we developed after the original 
manuscript was submitted.
We hope to publish a more detailed discussion elsewhere.

We discuss in detail the situation of two connected grains to demonstrate
again the essential physics. For two grains, the AES action can always be
expressed in terms of the average phase $\phi_{av}=(\phi_1+\phi_2)/2$
and relative phase $\phi=(\phi_1-\phi_2)$, where $\phi_1$ and $\phi_2$
are the phases on first and second grains.
\begin{eqnarray}
S=\frac{1}{4E_c} \int_0^\beta d\tau\left( \frac{d\phi_1}{d\tau} \right)^2+
\frac{1}{4E_c} \int_0^\beta d\tau \left( \frac{d\phi_2}{d\tau} \right)^2+
S_t (\phi_1-\phi_2)= \nonumber \\
=\frac{1}{4(E_c/2)} \int_0^\beta d\tau\left( \frac{d\phi_{av}}{d\tau} \right)^2+
\frac{1}{4(2E_c)} \int_0^\beta d\tau\left( \frac{d\phi}{d\tau} \right)^2+
S_t (\phi).
\end{eqnarray}
The part of the action for the average phase is trivial and is easily transformed
to the charge representation. In doing so it is necessary to satisfy carefully
correct Matsubara boundary conditions of the original fields $\phi_1$ and $\phi_2$. 

The crucial question is what the minimum
charging energy of two grains is. 
Is the minimum charging energy $E_c/2$ or still $E_c$ 
as for a single grain? For a two-grain system, this issue can be
addressed by considering the phase correlation function for one of the grains,
\begin{eqnarray}
C_{1}(\tau) & = & \langle \cos(\phi_{1}(\tau)-\phi_{1}(0)) \rangle
  \nonumber \\
  & = & \langle \cos(\phi_{av}(\tau)-\phi_{av}(0))\rangle\langle
         \cos((\phi(\tau)-\phi(0))/2)\rangle.
\end{eqnarray}
The part of the action for the average phase $\phi_{av}$ corresponds
to the charging energy $E_c/2$ for the total charge $(q_1+q_2)$, 
quantised and equal to one. From this part, $C_{1}(\tau)$ gets a
contribution $\exp(-E_c\tau/2)$. If $g$ were zero, then the relative
phase contribution to $C_{1}(\tau)$ is also $\exp(-E_c\tau/2)$, so
that $C_{1}(\tau)=\exp(-E_{c}\tau)$, in accordance with our
expectation for isolated grains. For a finite value of $g$,   
we demonstrate below that correlation functions of the relative phase
fluctuations at large $\tau$ decrease only algebraically as a function
of temperature, and do not show a hard Coulomb
gap\cite{altland}. The gaplessness of the relative phase fluctuations 
unambiguously proves that the minimum charging energy of the 
two grains is halved, $E_c/2$, and associated only with the average phase. 
If instead, the relative phase correlator were gapped, with some 
effective charging energy $E^\star_c$, then the correlation
function would decrease exponentially at long-$\tau$ with the 
corresponding charging energy. 
Therefore
the question about the effective charging energy is equivalent to the question
of considering the long-$\tau$ asymptotics (or equivalently low temperatures)
of the correlation function $C_{1}(\tau)$. The last statement is of a 
general character,
since the long-time (or low temperature) asymptotics always reveals
the lowest energy excitations (or configurations of the $\phi$-field) 
of the system.
Namely, if the charge gap exists, this will become evident as 
an exponential decay
of the corresponding correlation function at long times. In the
literature, the charge gap is occasionally related to the amplitude of
Coulomb blockade oscillations as a function of a gate voltage on a grain. In our
case, the charge gap is the cost of putting one excess charge on a
grain. These two definitions are not necessarily the same. For
calculating the conductivity of the granular system, our definition of
the charge gap is the appropriate one.

Let us consider the relative phase
correlator
\begin{equation}
C(\tau)=\langle \cos(\phi(\tau)-\phi(0))\rangle,
\label{checking}
\end{equation}
which has been extensively studied in the literature, so a comparison
with known results is possible. Besides,
$C(\tau)\leq \langle \cos((\phi(\tau)-\phi(0))/2\rangle,$ so if we can
show that the large $\tau$ behaviour of $C(\tau)$ does not have a hard
gap, then it is true for $\langle \cos((\phi(\tau)-\phi(0))/2\rangle$ too.
We claim (and we are not the first ones) that the correlation function 
$C(\tau)$ of the single-phase action
decays, in fact, as a power-law, $(T/T_{*})^2/\sin^{2}(\pi T\tau)$, 
at very large $\tau$ and 
{\it not exponentially}. Here $T_{*}$ is an energy scale exponentially
small in $g$. 
Thus the correlation function of relative phase fluctuations is
not gapped for long times. This is a crucial point because a temperature
dependence of $C(\tau)$ that is not exponentially small in temperature
at large $\tau$ invariably leads to a soft activation behaviour of
conductivity at low temperatures (see concluding remarks in this
section). 
Several arguments based on general results of statistical physics
and mesoscopics as well as our numerical Monte Carlo simulations prove
beyond doubt that the correlation function $C(\tau)$ decays 
algebraically (proportional to $1/\sin^{2}(\pi T\tau)$).
First, note that the action for relative phase fluctuations is a one-dimensional 
field theory with a long-range interaction, $gT^2/\sin^2(\pi T\tau)$, in 
imaginary time.
A general theorem of statistical physics due to 
Griffiths\cite{griffiths} states that the correlation function 
$C(\tau)$ cannot decay faster than
the interaction, $gT^2/\sin^2(\pi T\tau)$.
The exponential decay is much faster than algebraic decay and therefore not possible.
Second, it is widely recognised in the mesoscopics literature that at low temperatures
the tunnelling to a quantum dot is dominated by so-called inelastic (or elastic)
cotunnelling processes\cite{averin,zwerger}. In the case of two grains, cotunnelling
processes which are second order processes in the conductance $g$
correspond to creation of electron-hole pairs on both grains. These processes
are the lowest-energy gapless processes which can be closely associated
with long-$\tau$ behaviour of $C(\tau)$, once again this demonstrates
that the charge gap at low temperatures is effectively zero.
In fact, the picture of the charge sharing can be equally 
well discussed in terms
of the balance between Coulomb blockade and cotunnelling processes
for a finite set of grains. Our results concerning cotunnelling processes
are somewhat non-trivial, because we describe these processes 
in terms of the parameter $\gamma (T)$ for $g \gg 1$ 
(unlike the originally considered case of $g \ll 1$ of Ref.\cite{averin,zwerger}). 
Third, we undertook numerical simulations of the single-phase AES
action using the path integral Monte Carlo method. 
It is possible to calculate
directly by this method, without any approximations, not only the correlation function $C(\tau)$
but also the parameter $\gamma$ as a function of $g$ and $T.$
Numerical results show clearly that the correlation function behaves
as $(T/T_{*})^{2}/\sin^{2}(\pi T \tau)$ in the large-$\tau$
limit. Note that a rough estimate of $C(\tau)$ as $C(\tau) \sim [\sum_k
\cos(2\pi k T \tau)\exp(-S(k))]/\exp(-S(k))$ using
Eq.(\ref{twograinaction}) gives
$C(\tau)\sim\sinh^2(\pi\gamma/2)/[\sinh^{2}(\pi\gamma/2)+\sin^{2}(\pi
  T\tau)]$. For small values of $\gamma$, the gaussian approximation
is inaccurate, nevertheless the Lorentzian long time behaviour of $C(\tau)$
inferred from Eq.(\ref{twograinaction}) clearly anticipates
the $(T/T_{*})^{2}/\sin^{2}(\pi T\tau)$ result of exact numerical calculations, with $\gamma\sim
T/T_{*}\ll 1$ and large $\tau$. 

\begin{figure}[h]
\includegraphics{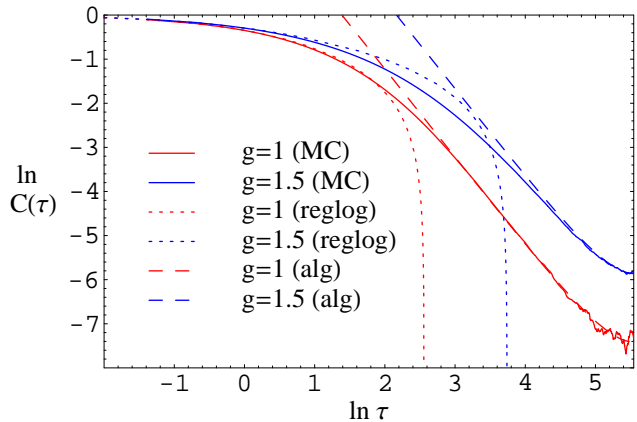}
\caption{\label{evidence_lnctau_lnbeta}  
$\ln C(\tau)$ vs $\ln \tau$ for $g=1$ (red) and $g=1.5$ (blue), at $\beta=768$.
The data (solid curves) show a crossover from logarithmic behaviour $C^\text{reglog}$(dotted curves) 
to the power-law behaviour $C^\text{alg}$(dashed).  Discrepancies at small $\tau$ are due to discretisation error.  
Note that $C^\text{reglog}$ is not exactly a power law (or a straight
line on a $log-log$ plot): 
the cosec squared flattens out at $\tau \sim \beta/2$.
}
\end{figure}

In what follows we summarise the results of the path integral Monte Carlo
simulations (the description of the method and further results will
be published elsewhere\cite{ylloh}).
In Fig.\ref{evidence_lnctau_lnbeta}  we present the correlation function $C(\tau)$ for $g=1$
and $g=1.5$. One can see
clearly that at short imaginary times-$\tau$, $C(\tau)=1-1/(\pi g)\ln(gE_c/2T)$,
which is consistent with $g_{ren}$ calculated in Eq.(\ref{grenormal1})
ignoring winding numbers and various  
others\cite{efetov1}. At long times-$\tau$, in
Fig.{\ref{evidence_lnctau_lnbeta}}, the correlation function 
decays as $1/\sin^{2}(\pi T\tau)$.

\begin{figure}[h]
\includegraphics{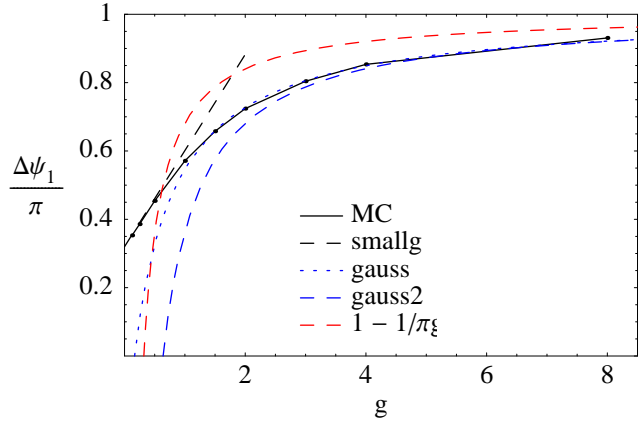}
\caption{\label{evidence_dpsi_g_beta16}  
$\Delta\psi_1(g)$, defined as the derivative with respect to $g$ 
of the normalised
  action (see Eq.(\ref{psi})), plotted against $g$ for $\beta=16$.
  Solid black line: Monte Carlo.  
Dashed black line: small-$g$ perturbation theory.  
Dotted blue line: $\Delta\psi^\text{gauss}_1$, using the gaussian
approximation of Eq.(\ref{gaussiangamma}).  
Dashed blue line: $\Delta\psi^\text{gauss2}_1$, using
Eq.(\ref{effectivegamma}) which can be deduced from
Eq.(\ref{gaussiangamma}) using Stirling's approximation.  
Dashed red line: $(1-1/\pi g)$, i.e., the derivative of $g_{ren}$ with
respect to $g$.   
}
\end{figure}

\begin{figure}[h]
\includegraphics[width=10.5cm]{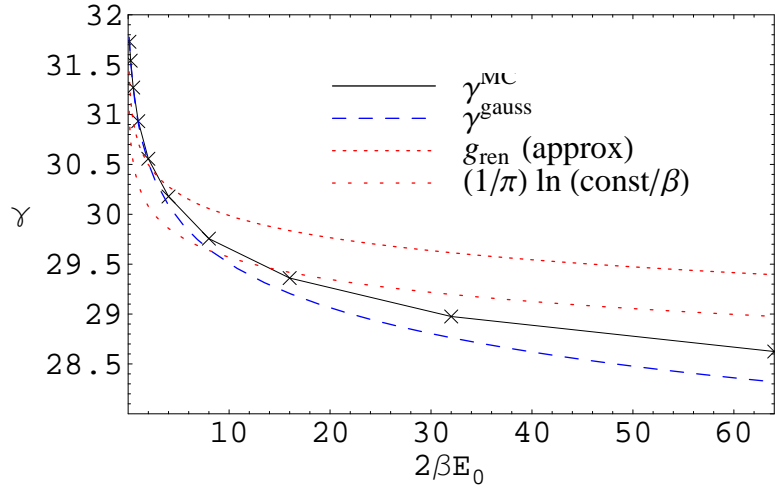}
\caption{\label{evidence_gamma_beta_g32}  
$\gamma$ versus $\beta$ for $g=32$. Solid black line: Monte Carlo
  calculation. Dashed blue line: $\gamma(T)$ in Eq.(\ref{gaussiangamma})
obtained by considering gaussian fluctuations. Top red line: 
$g_{ren}(T)$ extracted from
$C(\tau)$ with $\tau$ proportional to $\beta$. Bottom red line: 
$g_{ren}(T)$ with 
adjusted cutoff, $g_{ren}(T)=[g-(1/\pi)\ln(\text{const.}E_{c}/T)]$.
Clearly, $(1/\pi)\ln T$ is not in good agreement with 
Monte Carlo results for any adjustment of the cutoff. The gaussian
approximation gives a much better agreement with Monte Carlo
results.    
}
\end{figure}

The detailed behaviour of the parameter $\gamma$ is given in Fig.\ref{evidence_dpsi_g_beta16}
as a function of $g$ and in Fig.\ref{evidence_gamma_beta_g32}  as a function of temperature $T$.
The exact calculation of the determinant of the residual fluctuations
in the gaussian approximation gives instead of $\pi\gamma |k_{12}|$ in Eq.\ref{twograinaction}
the following expression:
\begin{eqnarray}
\pi\gamma|k|=\pi g|k|- \nonumber \\
			\ln \frac{ 
				\Gamma\left(1+k+\frac{x-\sqrt{x}\sqrt{x+4k}}{2}\right) ~
				\Gamma\left(1+k+\frac{x+\sqrt{x}\sqrt{x+4k}}{2}\right)
			}{\Gamma(1+k) ^2~ \Gamma(1+x)}, 
\label{gaussiangamma}
\end{eqnarray}	
where $x=\frac{gE_c}{\pi T}$ and for brevity we denote the relative winding
number $k_{12}\equiv k$ as $k$.
This expression is more precise than the 
expression (see Eq.\ref{effectivegamma})
calculated using Stirling's approximation. 

Fig.\ref{evidence_dpsi_g_beta16} compares the
derivative of the action, 
\begin{equation}
\Delta\Psi_1(g)=\partial [S(k=1,g)-S(k=0,g)]/\partial g,
\label{psi}
\end{equation}
evaluated numerically, with various analytic approximations. The blue
dotted curve is the gaussian approximation of
Eq.(\ref{gaussiangamma}) and the blue dashed curve is
Eq.(\ref{effectivegamma}) which can be shown to follow from
Eq.(\ref{gaussiangamma}) using Stirling's approximation for the gamma 
functions. The red dashed curve corresponds to setting
$\gamma(T)=g_{ren}$. 

Fig.\ref{evidence_gamma_beta_g32} compares $\gamma(T)$, calculated
numerically, with the gaussian approximation in
Eq.(\ref{gaussiangamma}) (blue dashed
curve), and $g_{ren}(T)$ calculated from $C(\tau),\,\tau\propto\beta,$
(top red curve) and $g_{ren}(T)$ with an adjustable cutoff (lower red curve), 
as a function of temperature.  

We thus observe that the renormalisations of the quantities $g_{ren}$
and $\gamma$ as a function of temperature are {\it different} for large $g \gg 1$.
Although this observation is not essential by itself for the charge sharing mechanism,
the numerical simulation shows directly that the renormalisation of $\gamma$ is stronger
(see below). 
The stronger renormalisation of $\gamma$ in comparison with $g_{ren}$ should presumably be
associated with nearly zero-modes of residual fluctuations which exist
around winding number trajectories.
 Namely, the gaussian
fluctuations are stronger around winding number trajectories than around
non-winding ($k=0$) trajectory because the square averaged fluctuations
for zero modes (for $n \leq k$) are much stronger  $<\phi^2_n>_{zm} =4E_c/(\pi T)$ than  $<\phi^2_n> \sim 1/g$
for simple gaussian residual fluctuations. Note that zero-mode fluctuations 
need to be considered beyond gaussian approximation for $E_c \gg T$,
because $<\phi^2_n>_{zm}$ becomes easily much larger than
$(2\pi)^2$. Therefore fluctuations beyond logarithmic 
renormalisations are naturally expected and do occur as seen numerically.

We end with two remarks. Since the temperature dependence of the relative
phase correlator at large $\tau$ is not an exponentially small function of
temperature, but only a power law, optimising the
probability of a charge shared among $N$ 
grains, $P_{N}\sim \gamma(T)^{N-1}\exp(-E_{c}/NT)\approx
\exp(-2\sqrt{E_{c}\ln(\gamma^{-1})/T})$, 
gives a temperature dependence of the exponent that is always weaker than Arrhenius'
law, $P\sim \exp(-E_{c}^{*}/T)$. 
Thus the temperature dependence of the optimum probability $P_{N}$ is 
the soft activation behaviour. Second, it has not escaped our attention
that even for $g<1$, inelastic cotunnelling processes should make the
charge sharing possible\cite{ylloh}.


\vskip 1cm
\begin{thebibliography}{10}
\bibitem{zaikin}
S.V. Panyukov, A.D. Zaikin, Phys.Rev.Lett. {\bf 67}, 3168 (1991).
\bibitem{grabert} G. Goppert, H. Grabert, Eur. Phys. J. B 16, 687 (2000).
\bibitem{efetov1}K. B. Efetov, and A. Tschersich, Europhysics Lett. \textbf{59}, 114
(2002). 
\bibitem{abeles1}B. Abeles \emph{et al.}, Adv. Phys. \textbf{24}, 407 (1975).
\bibitem{ambegaokar1}V. Ambegaokar \emph{et al.}, Phys. Rev. Lett. \textbf{48}, 1745 (1982);
for a review see G. Sch\"{o}n, and A. D. Zaikin, Phys. Rep. \textbf{198},
237 (1990). 
\bibitem{chui1}T. Chui \emph{et al.}, Phys. Rev. B \textbf{23}, 6172 (1981). 
\bibitem{simon1}R. W. Simon \emph{et al.}, Phys. Rev. B \textbf{36}, 1962 (1987). 
\bibitem{gerber1}A. Gerber \emph{et al.}, Phys. Rev. Lett. \textbf{78}, 4277 (1997). 
\bibitem{devoret1}M. H. Devoret \emph{et al.}, Phys. Rev. Lett. \textbf{64}, 1824 (1990). 
\bibitem{girvin1}S. M. Girvin \emph{et al.}, Phys. Rev. Lett. \textbf{64}, 3183 (1990). 
\bibitem{altshuler1}B. L. Altshuler, and A. G. Aronov, in \emph{Electron-Electron Interaction
in Disordered Systems}, edited by A. L. Efros, and M. Pollak, (North
Holland, Amsterdam, 1985). 
\bibitem{beloborodov2}I. S. Beloborodov \emph{et al.},
  Phys. Rev. Lett. {\bf 91}, 246801 (2003). 
\bibitem{beloborodov1}I. S. Beloborodov \emph{et al.}, Phys. Rev. B \textbf{63}, 115109
(2001). 
\bibitem{hybrid1} This incoherent hybridisation of charge density between
  grains has to be distinguished from wave-function hybridisation
  (linear combination) in coherent, non-dissipative quantum mechanical systems.
\bibitem{falci} G. Falci, G. Schon, and G.T. Zimanyi, Phys.Rev.Lett.
{\bf 74}, 3257 (1995).
\bibitem{mycomment}In Eq.(\ref{env1}) if we choose to normalise
  the partition function by the partition function for
  $(q_{1}-q_{2})=0$, the factor $\gamma$ in the numerator of the second term
  of the equation should be replaced with $\gamma^{2}$. This does not
  affect our physical picture of charge sharing.  
\bibitem{degennes}P. de Gennes, \emph{Scaling Concepts in Polymer Physics}, Cornell
University Press, London (1979). 
\bibitem{jaeger}H.M. Jaeger \emph{et al.} Phys. Rev. B \textbf{40}, 182 (1989).
\bibitem{efros1}A. L. Efros, and B. I. Shklovskii, J. Phys. C
  \textbf{8}, L49 (1975); J. Zhang and B. I. Shklovskii, {\tt
  cond-mat/0403703} (2004).
\bibitem{ando1}Y. Ando \emph{et al.}, Phys. Rev. Lett. \textbf{75}, 4662 (1995). 
\bibitem{boebinger1}G. S. Boebinger \emph{et al.}, Phys. Rev. Lett. \textbf{77}, 5417
(1996). 
\bibitem{lang1}K.M. Lang \emph{et al.}, Nature \textbf{415}, 412 (2002); S.H. Pan
\emph{et al.}, Nature \textbf{413}, 282 (2001). 
\bibitem{cheong1}S-W. Cheong \emph{et al.}, Phys. Rev. B \textbf{37},
  5916 (1988); 
B.~Ellman \emph{et al.}, Phys. Rev. B \textbf{39}, 9012 (1989). 
\bibitem{griffiths}R.B.Griffiths, J. Math. Phys. (N.Y.) \textbf{8},
  478 (1967); J.Ginibre, Commun. Math. Phys. \textbf{16}, 310 (1970).
\bibitem{altland} A.Altland, L.I.Glazman, and A.Kamenev,
  Phys. Rev. Lett. {\bf 92}, 026801 (2004).
This work proposed an insulating phase with hard activation gap 
at low temperatures. It seems that the approximation of non-interacting 
instantons used in the work\cite{altland}
(and in some others, e.g. Ref.\cite{zaikin}) may not be valid, as for
  instance has been discussed in Ref.\cite{falci}. 
\bibitem{averin}
D.V. Averin, Yu.V. Nazarov, Phys.Rev.Lett. {\bf 65}, 2446 (1990).
\bibitem{zwerger}W.Zwerger and M.Scharpf, Zeitschrift Phys. B
  \textbf{85}, 421 (1991).   
\bibitem{ylloh}
Y.L. Loh, V. Tripathi, and M. Turlakov, in preparation (2004).
\end{thebibliography}
\end{document}